\documentclass[aps,twocolumn,pra,showpacs,floatfix]{revtex4}
\usepackage{epsfig}
\usepackage{graphicx}
\usepackage{dcolumn}
\usepackage{amssymb,amsmath}
\usepackage{mathrsfs}
\begin{document}

\title{Relativistic frequency shifts in Cr, Ti, Fe, Ni, Ca, Na, and V to search for variation in the fine structure constant}

\author{V. A. Dzuba, V. V. Flambaum}

\affiliation{School of Physics, University of New South Wales, Sydney 2052, Australia}

\author{M. T. Murphy, D. A. Berke}

\affiliation{Centre for Astrophysics and Supercomputing, Swinburne University of Technology, Hawthorn, Victoria 3122, Australia}

\begin{abstract}

Sensitivity of the frequencies of twenty two atomic transitions to the variation of the fine structure constant $\alpha$ is calculated.
The findings are to be used in search for possible variation of $\alpha$ across our Galaxy using
high quality spectroscopic data for stars similar to our Sun.
\end{abstract}

%\pacs{31.15.A-,11.30.Er}

\maketitle

\section{Introduction}

Theories unifying particle physics and cosmology predict possible variation of the fundamental constants in space and time. Values of these constants may depend on a cosmological scalar field representing dark energy or dark matter~\cite{Uzan,Olive,Stadnik,Eichhorn,Davoudiasl}.

Many searches for variations in the low-energy value of the
fine-structure constant, $\alpha\equiv e^2/\hbar c$, rely on
understanding the sensitivity of atomic transtion frequencies on
$\alpha$ (e.g.\cite{DzuFlaWeb99,DzuFlaCJP}).
%[MTM: Vladimir or Victor, please add most appropriate
%references for this statement). 
Indeed, these were recently calculated
for 13 transitions seen in infrared stellar spectra and used to probe
for differences between $\alpha$ in giant stars near the Galactic
Centre and the laboratory value \cite{Hees_2020PhRvL.124h1101H}. This
is also required for a new probe of $\alpha$-variation across our
Galaxy, using stars very similar to our Sun, called `solar twins'
\cite{Murphy_22,Berke_22a,Berke_22b}. This star-to-star comparison
largely avoids systematic errors in the usual star-to-laboratory
approach, particularly effective shifts and asymmetries in absorption
lines from several physical processes as they arise in stellar
atmospheres.

The solar twins approach seeks to take advantage of the richness of
optical stellar spectra, with potentially hundreds of strong (but
unsaturated) absorption lines available, in principle, for comparison
between stars to measure any differences in $\alpha$. However, in
practice, there are many reasons to carefully select only transitions
which are likely to be most reliable \cite{Berke_22b}. In particular,
using closely-separated pairs of transitions, which have very similar
strengths in solar twin spectra, suppresses many systematic errors of
instrumental and astrophysical origin. To use the thousands of
high-quality archival stellar spectra available from the High-Accuracy
Radial velocity Planetary Searcher (HARPS) spectrograph
\cite{Mayor_2003} on the European Southern Observatory's
(ESO) 3.6-m telescope at La Silla Observatory (Chile), 164 different
transitions were selected, which form 229 close pairs
\cite{Berke_22b}. However, the sensitivity of these transitions to
variations in $\alpha$ have not been calculated before. Here we
calculate sensitivies for 22 of these transitions, forming 17 pairs,
which have been used to compare $\alpha$ between 18 solar twins in
\cite{Murphy_22,Berke_22a}.

\section{Calculations}

The dependence of atomic transition frequencies on the fine structure constant $\alpha$ was studied before for a number of systems (see, e.g. \cite{DzuFlaWeb99,DzuFlaCJP}).
In particular, Ti and Fe where considered in Refs.~\cite{BerEtAl04,DzuFla08}. However, in present work we use atomic transition which were never considered before.
To perform the calculations we use similar approaches as before accompanied with a new method for atoms with complicated electronic structure (many electrons in open shells). {For atomic transitions it is convenient to use atomic units $\hbar=m_e=|e|=1$. In these units dependence of electron  energies and transition frequencies on $\alpha$ appears due to the relativistic corrections.
It is convenient to present frequencies of atomic transitions in a form}
\begin{equation}
\omega(x) = \omega_0+qx,
\label{e:wq}
\end{equation}
where $x=(\alpha/\alpha_0)^2-1$, $\omega_0$ and $\alpha_0$ are the present values of the transition frequency and the fine structure constant, $q$ is sensitivity coefficient which can be found by varying the value of $\alpha$ in computer codes
\begin{equation}
q = \frac{\omega(\Delta x)-\omega(-\Delta x)}{2\Delta x}.
\label{e:q}
\end{equation}
We use small value of $\Delta x$ ($\Delta x =0.01$) to avoid non-linear effects, e.g. those which are caused by level crossing~\cite{DzuEtAl02}.

To calculate transition frequencies, we use a range of methods depending on the number of electrons above closed shells. 
For atoms with one external electron (e.g., Na~I) we use the correlation potential method~\cite{DzuFlaSilSus87}.
In this method the calculations start from the relativistic Hartree-Fock (RHF) procedure for the closed-shell core, with one external electron removed.
Then, states of external electron are calculated by solving RHF-like equations which also include core-valence correlations.
\begin{equation}
(\hat H_0 + \hat \Sigma_1 - \epsilon_v)\psi_v = 0.
\label{e:BO}
\end{equation}
Here $\hat H_0$ is the RHF Hamiltonian, $\hat \Sigma_1$ is a single-electron non-local operator (correlation potential) which describes correlation interaction of an external electron with the core. We calculate $\hat \Sigma_1$ in the lowest, second order of the many-body perturbation theory using the B-spline technique~\cite{B-spline}. 
Index $v$ in (\ref{e:BO}) numerates  states of the valence electron. Solutions of (\ref{e:BO}) are usually called Bruekner orbitals (BO).
Transition frequencies are found as a difference between two BO energies, $\omega_{vw} = \epsilon_v-\epsilon_w$. Calculated energies of the states of Na~I which are used in the analysis are compared with experimental data in Table~\ref{t:Te}.

\begin{table}
\caption{\label{t:Te}
Energy levels used in the analysis, comparison of the theory with experiment (cm$^{-1}$).}
\begin{ruledtabular}
\begin{tabular}{llllrr}
\multicolumn{1}{c}{$Z$}&
\multicolumn{1}{c}{Atom/}&
\multicolumn{2}{c}{State} &
%\multicolumn{1}{c}{$J$} &
\multicolumn{2}{c}{Energy} \\
&\multicolumn{1}{c}{Ion}&&&
\multicolumn{1}{c}{Expt.~\cite{NIST}}&
\multicolumn{1}{c}{Theory} \\
\hline
11& Na I & $2p^63p     $                  &   $^2$P$^{\rm o}_{1/2}$& 16956.170 &  16862 \\
      &  & $2p^63p     $                  &   $^2$P$^{\rm o}_{3/2}$& 16973.366 &  16881 \\
      && $2p^65s           $            &   $^2$S$_{1/2}$& 33200.673 &  33057 \\ 
      &&                                &                 &                & \\
20& Ca I & $3p^64s4p   $                  &   $^3$P$^{\rm o}_1$  & 15210.063  &  15544 \\
       & & $3p^63d4s   $                  &   $^3$D$_1$  &  20335.360 &   19568 \\
      && $3p^64s5s   $                  &   $^3$S$_1$  &  31539.495 & 31816 \\
      && $3p^64s5p   $                  &   $^3$P$^{\rm o}_0$  &  36547.688 & 36766 \\
      &&                                &                 &                & \\
22& Ti I & $3d^3(^2$H$)4s $               & a $^3$H$_5$  &  18141.265 & 21534 \\
      && $3d^3(^2$H$)4s $               & a $^3$H$_6$  &  18192.570 & 21369 \\
      && $3d^3(^2$H$)4p $               & x $^3$H$^{\rm o}_5$  &  39152.103 & 41034 \\
      && $3d^3(^2$H$)4p $               & x $^3$H$^{\rm o}_6$  &  39198.323 &  41050 \\

 &Ti II& $3d^2(^1$D$)4s $               & a $^2$D$_{3/2}$&  8710.567 & 10577 \\
      && $3d^2(^1$G$)4s $               & b $^2$G$_{7/2}$& 15265.700 &  17936 \\
      && $3d^2(^3$F$)4p $               & z $^4$G$^{\rm o}_{5/2}$&  29544.454 & 29005  \\
      && $3d^2(^3$F$)4p $               & z $^2$G$^{\rm o}_{7/2}$& 34543.380 &  33930 \\
      &&                                &                 &                & \\
24& Cr I & $3d^44s^2   $                  & a $^5$D$_2$  &  7927.441  & 7114 \\
      && $3d^44s^2   $                   & a $^5$D$_3$  &  8095.184  & 7161 \\
      && $3d^4(^5$D$)4s4p(^3$P$^{\rm o})$& y $^5$P$^{\rm o}_1$  &  29420.864 & 29204 \\
      && $3d^4(^5$D$)4s4p(^3$P$^{\rm o})$& y $^5$P$^{\rm o}_2$  &  29584.571 &  29284 \\
      &&                                &                 &               & \\
23& V I  & $3d^4(^5$D$)4s $               & a $^6$D$_{9/2}$&  2424.809 & 2753 \\
      && $3d^3(^4$F$)4s4p(^3$P$^{\rm o})$& z $^6$D$^{\rm o}_{9/2}$&  18438.044 & 15575  \\
      &&                                &                 &                & \\
26& Fe I & $3d^7(^4$F$)4s $               & a $^3$F$_2$  & 12968.554 & 11964 \\ 
 &even & $3d^7(^4$P$)4s $               & a $^5$P$_3$  & 17550.181 & 23061 \\
      && $3d^7(^4$P$)4s $               & a $^5$P$_1$  & 17927.382 & 22558 \\
      && $3d^64s^2   $                  & a $^3$P$_1$  & 19552.478 & 24345 \\
      && $3d^64s^2   $                  & a $^3$H$_4$  & 19788.251 & 22233 \\
      && $3d^64s^2   $                  & b $^3$F$_3$  & 20874.482 & 25230 \\
      && $3d^64s^2   $                  & b $^3$F$_2$  & 21038.987 & 25746 \\

 &Fe I & $3d^7(^4$F$)4p       $         & y $^5$D$^{\rm o}_2$  &  33801.572 & 33476 \\
 &odd  & $3d^6(^5$D$)4s4p(^3$P$^{\rm o})$ & z $^3$P$^{\rm o}_2$  &  33946.933 & 33572 \\
     & & $3d^7(^4$F$)4p       $         & y $^5$D$^{\rm o}_0$  &  34121.603 & 34050 \\
      && $3d^7(^4$F$)4p       $         & z $^3$G$^{\rm o}_3$  &  36079.372 & 35648 \\
      && $3d^6(^5$D$)4s4p(^1$P$^{\rm o})$ & y $^5$P$^{\rm o}_2$  &  37157.566 & 39130 \\
      && $3d^7(^4$F$)4p       $         & y $^3$F$^{\rm o}_3$  &  37162.746 & 37388 \\
      && $3d^7(^4$F$)4p       $         & y $^3$D$^{\rm o}_1$  &  38995.736 & 39402 \\
      &&                                &                 &                & \\
28& Ni I & $3d^8(^1$D$)4s^2$              &  $^1$D$_2$  &  13521.347 & 15768 \\
      && $3d^9(^2$D$)4p       $         &   $^1$P$^{\rm o}_1$   &  32982.260 & 30508 \\ 
      && $3d^9(^2$D$)4p       $         &   $^3$D$^{\rm o}_3$  &  29668.893 & 28002 \\
\end{tabular}
\end{ruledtabular}
\end{table}

For atoms with two external electrons (e.g., Ca~I) we use a combination of the configuration interaction and many-body perturbation theory (CI+MBPT, \cite{DzuFlaKoz96}). Calculations are performed in the $V^{N-2}$ approximation~\cite{Dzu05}. The self-consistent Hartree-Fock procedure is done for the double ionised ion. Single-electron basis states for valence electrons are calculated in the field of the ion using the B-spline technique similar to how it is done for a single-valence-electron atoms. The effective CI Hamiltonian for two valence electrons has the form
\begin{equation}\label{e:CI}
\hat H^{\rm CI} = \hat h_1(r_1) + \hat h_1 (r_2) + \hat h_2(r_1,r_2).
\end{equation}
Here $\hat h_1 = \hat H_0 + \hat \Sigma_1$ is a single-electron part of the Hamiltonian. 
It includes the RHF operator $\hat H_0$ and the correlation potential $\hat \Sigma_1$, the same as in (\ref{e:BO}). The two-electron part of the CI Hamiltonian is the sum of the Coulomb interaction and correlation operator $\Sigma_2$, $\hat h_2(r_1,r_2) = e^2/|r_1-r_2| + \hat \Sigma_2(r_1,r_2)$. 
The later can be understood as screening of the Coulomb interaction between valence electrons by core electrons (see Ref.~\cite{DzuFlaKoz96} for details).
Calculated energies of the states of Ca~I which are used in the analysis are compared with experimental data in Table~\ref{t:Te}.

\begin{table*}
\caption{\label{t:Tq}
Electric dipole transitions used in the analysis and their coefficients of sensitivity to the variation of the fine structure constant.}
\begin{ruledtabular}
\begin{tabular}{lllcr llcr c}
\multicolumn{1}{c}{Atom/}&
\multicolumn{2}{c}{Lower state} &
\multicolumn{1}{c}{$J$} &
\multicolumn{1}{c}{Energy} &
\multicolumn{2}{c}{Upper state} &
\multicolumn{1}{c}{$J$} &
\multicolumn{1}{c}{Energy} &
\multicolumn{1}{c}{$q$} \\
\multicolumn{1}{c}{Ion}&&&&
\multicolumn{1}{c}{[cm$^{-1}$]}&&&&
\multicolumn{1}{c}{[cm$^{-1}$]}&
\multicolumn{1}{c}{[cm$^{-1}$]} \\

\hline
Na I & $2p^63p     $&   $^2$P$^{\rm o}$& 1/2& 16956.170 & $2p^65s           $&   $^2$S & 1/2&  33200.673 &    7(1)   \\
Na I & $2p^63p     $&   $^2$P$^{\rm o}$& 3/2& 16973.366 & $2p^65s           $&   $^2$S & 1/2&  33200.673 &  -11(1)   \\
      &              &         &    &           &                    &         &    &            &           \\
Ca I & $3p^64s4p   $&   $^3$P$^{\rm o}$& 1  & 15210.063 & $3p^64s5s         $&   $^3$S & 1  &  31539.495 &  -69(7)   \\
Ca I & $3p^63d4s   $&   $^3$D & 1  & 20335.360 & $3p^64s5p         $&   $^3$P$^{\rm o}$& 0  &  36547.688 & -475(24)  \\
      &              &         &    &           &                    &         &    &            &           \\
Ti I & $3d^3(^2$H$)4s $& a $^3$H & 5  & 18141.265 & $3d^3(^2$H$)4p       $& x $^3$H$^{\rm o}$& 5  &  39152.103 &  440(60)  \\
Ti I & $3d^3(^2$H$)4s $& a $^3$H & 6  & 18192.570 & $3d^3(^2$H$)4p       $& x $^3$H$^{\rm o}$& 6  &  39198.323 &  460(60)  \\
      &              &         &    &           &                    &         &    &            &           \\
Ti II& $3d^2(^1$D$)4s $& a $^2$D & 3/2&  8710.567 & $3d^2(^3$F$)4p       $& z $^4$G$^{\rm o}$& 5/2&  29544.454 &   90(90)  \\
Ti II& $3d^2(^1$G$)4s $& b $^2$G & 7/2& 15265.700 & $3d^2(^3$F$)4p       $& z $^2$G$^{\rm o}$& 7/2&  34543.380 &  340(150)  \\
      &              &         &    &           &                    &         &    &            &           \\
Cr I & $3d^44s^2   $& a $^5$D & 2  &  7927.441 & $3d^4(^5$D$)4s4p(^3$P$^{\rm o})$& y $^5$P$^{\rm o}$& 1  &  29420.864 & 490(60) \\
Cr I & $3d^44s^2   $& a $^5$D & 3  &  8095.184 & $3d^4(^5$D$)4s4p(^3$P$^{\rm o})$& y $^5$P$^{\rm o}$& 2  &  29584.571 &  490(60)  \\
      &              &         &    &           &                    &         &    &            &           \\
V I  & $3d^4(^5$D$)4s $& a $^6$D & 9/2&  2424.809 & $3d^3(^4$F$)4s4p(^3$P$^{\rm o})$& z $^6$D$^{\rm o}$& 9/2&  18438.044 & -600(60)  \\
      &              &         &    &           &                    &         &    &            &           \\
Fe I & $3d^7(^4$F$)4s $& a $^3$F & 2  & 12968.554 & $3d^7(^4$F$)4p       $& y $^5$D$^{\rm o}$& 2  &  33801.572 & -830(400) \\ %-830(80) 
Fe I & $3d^64s^2   $& a $^3$P & 1  & 19552.478 & $3d^7(^4$F$)4p       $& y $^3$D$^{\rm o}$& 1  &  38995.736 & 2840(300) \\
Fe I & $3d^7(^4$P$)4s $& a $^5$P & 1  & 17927.382 & $3d^6(^5$D$)4s4p(^1$P$^{\rm o})$& y $^5$P$^{\rm o}$& 2  &  37157.566 & -1500(900)  \\
Fe I & $3d^64s^2   $& a $^3$H & 4  & 19788.251 & $3d^7(^4$F$)4p       $& z $^3$G$^{\rm o}$& 3  &  36079.372 & 2710(500) \\
Fe I & $3d^64s^2   $& b $^3$F & 3  & 20874.482 & $3d^7(^4$F$)4p       $& y $^3$F$^{\rm o}$& 3  &  37162.746 & 3010(500) \\
Fe I & $3d^7(^4$P$)4s $& a $^5$P & 3  & 17550.181 & $3d^7(^4$F$)4p       $& y $^5$D$^{\rm o}$& 2  &  33801.572 &  340(400)  \\
Fe I & $3d^7(^4$P$)4s $& a $^5$P & 1  & 17927.382 & $3d^7(^4$F$)4p       $& y $^5$D$^{\rm o}$& 0  &  34121.603 &  430(500)  \\
Fe I & $3d^64s^2   $& b $^3$F & 2  & 21038.987 & $3d^7(^4$F$)4p       $& y $^3$F$^{\rm o}$& 3  &  37162.746 & 2870(500) \\
Fe I & $3d^7(^4$P$)4s $& a $^5$P & 1  & 17927.382 & $3d^6(^5$D$)4s4p(^3$P$^{\rm o})$& z $^3$P$^{\rm o}$& 2  &  33946.933 & -1300(900)  \\ 
      &              &         &    &           &                    &         &    &            &           \\
Ni I & $3d^8(^1$D$)4s^2$&  $^1$D & 2  & 13521.347 & $3d^9(^2$D$)4p       $&   $^1$P$^{\rm o}$& 1  &  32982.260 & 4950(500) \\
Ni I & $3d^8(^1$D$)4s^2$&  $^1$D & 2  & 13521.347 & $3d^9(^2$D$)4p       $&   $^3$D & 3  &  29668.893 & 5500(550) \\
 \end{tabular}
%\footnotetext[1]{Current clock state.}
\end{ruledtabular}
\end{table*}

For atoms with more than two external electrons (Ti~I, Ti~II, Cr~I, V~I, Fe~I, Ni~I) we use a combination of the configuration interaction method with the perturbation theory (the CIPT method~\cite{CIPT}). All electrons above the Ar-like core are considered as external electrons and treated with configuration interaction (CI) approach. The CI Hamiltonian has a form
\begin{equation}\label{e:CIn}
\hat H^{\rm CI} = \sum_i^{N_v}\hat H_0(r_i) + \sum_{i<j}^{N_v}e^2/r_{ij},
\end{equation}
where $\hat H_0$ is the RHF operator, second term presents Coulomb interaction between external electrons, the summation goes over all external (valence) electrons $N_v$. Correlation operators $\Sigma_1$ and $\Sigma_2$ are not included for reasons explained below.
The number of valence electrons varies from $N_v=4$ for Ti to  $N_v=10$ for Ni. With the number of external electrons being that large, the number of lines in the CI matrix may go up to $\sim 10^8$ in the standard CI approach. This would make the calculations impossible. 
Therefore, we use the CIPT~\cite{CIPT} approach, in which the off-diagonal matrix elements between high-energy states are neglected. Indeed, it is easy to show that these matrix elements formally appear in a higher order of the perturbation theory and are suppressed by large energy denominators. 
This allows to reduce the size of the effective CI matrix by many orders of magnitude. The high-energy states are still included in the correction to the CI matrix elements between low-energy states,
\begin{equation}\label{e:CIPT}
\langle a |\hat H^{\rm eff} |b\rangle = \langle a |\hat H^{\rm CI} |b\rangle + \sum_n \frac{\langle a |\hat H^{\rm CI} |n\rangle\langle n |\hat H^{\rm CI} |b\rangle}{E-E_n}.
\end{equation}
Here $\hat H^{\rm eff}$ is the effective CI operator, $|a\rangle$ and $|b\rangle$ are many-electron single-determinant low-energy basis states, $|n\rangle$ are high-energy many-electron basis states, $E_n$ is the energy of the high-energy state, $E_n=\langle n|\hat H^{\rm CI}|n\rangle$, $E$ is the energy of the state of interest. 
Since the later energy is not known, the CIPT equations, ($H^{\rm eff} -EI)X=0$, are solved iteratively, using the energy found on the previous iteration to calculate second term in (\ref{e:CIPT}), (see Ref.~\cite{CIPT} for details).

Many-electron basis states are constructed using a set of single-electron basis states calculated in the $V^{N-1}$ approximation using the B-spline technique~\cite{B-spline}. The $V^{N-1}$ approximation means that the initial RHF procedure is done for a system with one external electron removed. 
In all considered systems we remove one $4s$ electron from the ground state and perform the initial RHF calculations for the [Ar]$3d^n4s$ configuration 
(the [Ar]$3d^3$ configuration for Ti).
The basis states above the core are calculated in the field of frozen core. These states are linear combinations of B-splines which are the eigenstates of the RHF Hamiltonian with the $V^{N-1}$ potential. In the CIPT method it is important to have good initial approximation so that high states can be treated perturbatively.
Therefore, we cannot include correlations between valence electrons and electrons in the Ar-like core via the $\Sigma_1$ and $\Sigma_2$ operators as it is done for a two-electron system, Eq.~(\ref{e:CI}). These operators are well defined only in the The $V^{N-M}$ approximation~\cite{Dzu05}, where $M$ is the number of valence electrons. In any other starting potential one has to include subtraction diagrams~\cite{DzuFlaKoz96} and even three-body diagrams~\cite{3body}, which is significant complication to the method. On the other hand, in the considered systems, the correlations are dominated by correlations between valence electrons which are treated pretty accurately in the CI technique. The effect of neglecting the core-valence correlations is smaller than the effect of neglecting the off-diagonal matrix elements between the high-energy states.

Calculated energies of the states of Ti~I, Ti~II, Cr~I, V~I, Fe~I and Ni~I, which are used in the analysis are compared with experimental data in Table~\ref{t:Te}.
The accuracy is lower than for systems with simple electron structure, like Na~I and Ca~II. However, it is sufficient for the purposes of the present work.

The pairs of transitions, which are used in the analysis, are presented in Table~\ref{t:Tq} together with the calculated sensitivity coefficients $q$.
Error bars for the $q$-coefficients are estimated by varying computational parameters, such as the value of $\Delta x$ in Eq.~(\ref{e:q}), 
the number of configurations included into the low-energy basis, etc. In most of cases the uncertainty is significantly smaller that the value of $q$.

\acknowledgments

V.A.D.\ and V.V.F.\ acknowledge the Australian Research Council for
support through grants DP190100974 and DP200100150, and M.T.M.\
acknowledges the Council's support through \textsl{Future Fellowship}
grant FT180100194.

%\bibliographystyle{apsrev}
%\bibliography{michael,dzuba}

%\end{document}

\end{document}